\documentclass[3p,times,twocolumn]{elsarticle}

\usepackage{ecrc}

\volume{00}

\firstpage{1}

\journalname{Nuclear Physics B Proceedings Supplement}

\runauth{Miguel-Angel Sanchis-Lozano and Edward Sarkisyan-Grinbaum}

\jid{nuphbp}

\jnltitlelogo{Nuclear Physics B Proceedings Supplement}

\usepackage{amssymb}
\usepackage[figuresright]{rotating}

\begin{document}

\begin{frontmatter}



\dochead{}

\title{Searching for hidden sectors in multiparticle production at the LHC}


\author[label1]{Miguel-Angel Sanchis-Lozano}
\author[label2,label3]{Edward Sarkisyan-Grinbaum}

\address[label1]{Department of Theoretical Physics and  IFIC, CSIC-University of Valencia, 46100 Burjassot}
\address[label2]{Department of Physics, The University of Texas at Arlington, Arlington TX 76019, USA}
\address[label3]{Department of Physics, CERN 1211 Geneva 23, Switzerland} 
\begin{abstract}
Most signatures of new physics in colliders have been studied so far on the
transverse plane with respect to the beam direction. 
In this work however we study the impact of a hidden sector beyond the Standard Model (SM)
on inclusive (pseudo)rapidity correlations and
moments of the multiplicity distributions, with special emphasis in the
LHC results.
\end{abstract}


\end{frontmatter}

% \linenumbers

\section{Introduction}
\label{}

Most signatures and signals 
of new phenomena are expected to be found at colliders 
in hard events on the transverse plane with respect to
the beam direction (i.e. at high transverse momentum $p_{\bot}$), 
where background is much reduced. 

Conversely, in this work we focus on rather diffuse soft
signals in $pp$ interactions at the LHC. For example, 
a non-standard state of matter from a Hidden Sector (HS)
would alter observables related to (pseudo)rapidity particle correlations and factorial moments
of multiplicity distributions \cite{SanchisLozano:2008te}.

\subsection{Hidden Sectors}

Among possible HS scenarios, hidden valley models are extensions 
of the SM when a new gauge group is added to the theory, 
leading to new bound states (v-particles) with relatively low masses 
for some values of the model parameters. Such scenarios
\cite{Strassler:2006im} are physically motivated by top-down models including
string theory constructions. Their experimental consequences and signatures for LHC 
experiments have been already studied \cite{Strassler:2008}. For example, v-hadrons might promptly 
decay back to SM fermions contributing to the parton shower hadronizing
to final-state particles.

\section{Multi-step cascade: HS as an extra contribution}

Following the old ideas of multiperipheral model \cite{Dremin:1977wc,Kittel}, we will assume that 
particles are produced by {\em clusters} from the interaction of
two active partons of the colliding protons yielding a fragmenting string.
We treat this possibility as resulting from a 2-step process and apply
the same expression as an independent superposition of sources \cite{Dremin:2004ts}:
\begin{equation}\label{eq:ippi-2}
P^{(2)}(n)\ =\ \sum_{N_c}\ P(N_c)\ \sum_{n_i}\ \prod_{i=1}^{N_c}\ P^{(1)}(n_i)
\end{equation}
where $n$ and $N_c$ denote the number of (charged) particles and clusters, respectively;
$P^{(k)}(n)$ stands for a probability distribution where the superscript denotes a 
$k$-step cascade, such that $\sum_{n=1}^{\infty}P^{(k)}(n)=1$.

However, more than two 
partons can interact and more than one string can be produced per event, 
with a probability distribution $P(N_s)$ depending 
on the nature of the interacting bodies and cms 
energy. We consider this possibility as deriving from a 3-step process:
\begin{equation}\label{eq:ippi-3}
P^{(3)}(n)\ =\ \sum_{N_s}\ P(N_s)\ \sum_{n_j}\ \prod_{j=1}^{N_s}\ P^{(2)}(n_j)
\end{equation}
where $N_s$ denotes the number of strings per collision.

Finally, we consider that a new stage in the parton cascade  
can be produced as a consequence of a HS on top of the parton shower, as motivated
in the Introduction. We shall
refer to this scenario as a 
4-step process:
\begin{equation}\label{eq:ippi-4}
P^{(4)}(n)\ =\ \sum_{N_h}\ P(N_h)\ \sum_{n_k}\ \prod_{k=1}^{N_h}\ P^{(3)}(n_k)
\end{equation}
where $N_h$ denotes the number of hidden sources per collision and $P(N_h)$ its
probability distribution.

\section{Correlation functions}

\subsection{Two-particle rapidity correlations}

Rapidity inclusive 2-particle correlation function 
for inelastic collisions are defined as 
\begin{equation}\label{eq:C2}
C_2(y_1,y_2)= \rho_2(y_1,y_2)-\rho(y_1)\ \rho(y_2)
\end{equation}
where $y_i$ denotes the longitudinal rapidity of particle $i$.

Above we have introduced the single and 2-particle
(charged particle) rapidity densities:
\[
\rho(y)=\frac{1}{\sigma_{in}}
\int d^2p_{\bot}\frac{d^3\sigma}{dyd^2p_{\bot}}
\]
\[
\rho_2(y_1,y_2)=\frac{1}{\sigma_{in}} 
\int d^2p_{\bot 1}d^2p_{\bot 2} 
\frac{d^6\sigma}{dy_1d^2p_{\bot 1}dy_2d^2p_{\bot 2}}
\]
where $\sigma_{in}$ refers to the inelastic cross section;
the normalizations are obtained by integration over the selected
rapidity range.
\[
\int dy\ \rho(y)=  
\langle n \rangle\ ;  
\int  dy_1 dy_2\ \rho_2(y_1,y_2) =  
\langle n(n-1) \rangle
\]
\begin{equation}\label{eq:C2int} 
\int dy_1 dy_2\ C_2(y_1,y_2)= 
\ D^2-\langle n \rangle
\end{equation}
where $D^2=\langle n^2 \rangle - \langle n \rangle^2$
is the variance of the charged emitted particles. 
For a Poisson distribution $D^2=\langle n \rangle$, 
or equivalently the integral (\ref{eq:C2int}) is 
equal to zero corresponding to independent emission.

Quite generally, the inclusive correlation function is
split into two terms:
\begin{equation}{\label{eq:shortlongcorr}}
C_2(y_1,y_2)=C_2^{LR}(y_1,y_2)+C_2^{SR}(y_1,y_2)
\end{equation}
where the {\em short-range} (SR) term $C_2^{SR}$ is generally
assumed to be more sensitive to dynamical correlations, while
$C_2^{LR}$ stands for {\em long-range} (LR) correlations, usually
arising from mixing different topologies in events \cite{Capella:1978rg}.

\section{Factorial Moments in multiplicity distributions}

Factorial moments of multiplicity distributions  \cite{Dremin:2000ep,DeWolf:1995pc} are obtained from integration of the 
inclusive correlation functions. Normalized factorial moment of rank $q$ can be defined as:
\begin{equation}\label{eq:Fq}
F_q = \frac{\langle n(n-1)\cdots (n-q+1) \rangle}{\langle n \rangle^q}\ ;\ \ \ q=2,3,\cdots 
\end{equation}

Such factorial moments represent any correlation between the emitted particles in events. 
Cumulant $K_q$ represent
genuine $q$-particle correlations not reducible to the product of lower order correlations. They are 
defined through
\begin{equation}\label{eq:Kq}
K_q = \sum_{m=0}^{q-1}C_{q-1}^mK_{q-m}F_m
\end{equation}
where $C_{q-1}^m=1/mB(q,m)$ and $B(q,m)$ denotes the beta function.
 
It is also convenient to define the ratio 
\begin{equation}\label{eq:Hq}
H_q=\frac{H_q}{F_q}
\end{equation}
which appear in a natural manner as solutions of QCD equations for the generating functions of 
multiplicity distributions \cite{Dremin:1993sd}. QCD predicts that $H_q$ moments should oscillate 
(change of sign) as a 
function of the rank $q$ in $e^+e^-$ annihilations. This prediction has been confirmed experimentally, also for
$pp$ and heavy ion collisions. In the latter cases, such oscillations are ascribed to a multicomponent structrure
of the multiparticle production process (see \cite{Dremin:2000ep} for a review and references therein). 
This interpretation is quite relevant in our study since a new (HS) component could show up in
the production mechanism. 

\subsection{Cluster concept in hadronic production}

According to, e.g., Ref.\cite{Giovannini:1985mz},
a cluster if formed of all particles originating directly or indirectly
from a (e.g. quark or gluon) common $\lq\lq$ancestor''. 

Equations expressing the factorial moments as functions of parameters of the multiplicity distributions
of clusters, strings (and eventually hidden sources) are modified as more steps are introduced into the 
description of the parton cascade. In particular, we will denote by a superindex in 
the factorial moments, $F_q^{(j)}$, $K_q^{(j)}$ and $H_q^{(j)}$ with $j=2,3,4$, 
for two-, three- and four-step cascades, corresponding to
a single-string, multi-string and hidden-source scenario, respectively. 

Of course, as
more steps become included in description of the parton cascade, more parameters (encoding the complexity of the 
underlying soft hadronic dynamics) need to be introduced (see appendix). Nevertheless, 
the situation improves dramatically once a special type of distribution (for each step) is considered
(e.g. Poisson: $F_q=1$, $\forall q$).

\subsection{Negative Binomial Distribution}

The negative binomial distribution (NBD) deserves a special attention since it
has been used to describe quite successfully 
experimental multiplicities in a wide variety of proceses
and over a large energy range. The NBD is given by
\begin{equation}
P_n=\frac{\Gamma(n+k)}{\Gamma(n+1)\Gamma(k)}\biggl(\frac{\langle n \rangle}{k}\biggr)^n
\biggr(1+\frac{\langle n \rangle}{k}\biggr)^{-n-k}
\end{equation}
where $k$ is a parameter with the physical meaning of the number of independent sources.
Moreover
\begin{equation} 
\frac{1}{k}=\frac{D^2-\langle n\rangle}{\langle n \rangle^2} 
\end{equation}
The Poisson distribution is obtained in the limit $k\to \infty$.

\section{Two-step cascade: single string}

Let us first assume in the simplest (2-step) scenario that clusters are produced 
from a single string per collision, 
yielding final state particles (mainly hadrons).

\subsection{Two-particle correlations}

Let us introduce the rapidity 2-cluster 
correlation function 
\begin{equation}\label{eq:C2c}
C_2^{(c)}(y_1,y_2)= \rho_2^{(c)}(y_1,y_2)-\rho^{(c)}(y_1)\ \rho^{(c)}(y_2)
\end{equation}
with $\rho_2^{(c)}(y_1,y_2)$ and $\rho^{(c)}(y)$ refering to cluster densities
in rapidity space, satisfying the usual normalization condition
\begin{equation}
\int dy_{c1}dy_{c2}\ C_2^{(c)}(y_{c1},y_{c2})=D_c^2 - \langle N_c \rangle 
\end{equation}
where $D_c^2=\langle N_c^2 \rangle - \langle N_c \rangle^2$ 
denotes here the dispersion of the cluster distribution per collision.
 
The 2-particle correlation function
in Eq.(\ref{eq:C2}) can be split into two pieces
following Eq.(\ref{eq:shortlongcorr}):
\begin{equation}\label{eq:C2sl}
C_2(y_1,y_2)
= C_2^{LR}(y_1,y_2)\ +\ \langle N_c \rangle\ C_2^{(1)}(y_1,y_2)  
\end{equation}
where the piece $C_2^{(1)}$ corresponding to 
2-particle correlations $\lq\lq$inside'' a {\em single} cluster, reads
\begin{equation}\label{eq:single}
C_2^{(1)}(y_1,y_2)=
\rho_2^{(1)}(y_1,y_2)-\rho^{(1)}(y_1)\ \rho^{(1)}(y_2) 
\end{equation}
Note the assumption that $C_s^{(1)}(y_1,y_2)$ has no
explicit dependence on cluster rapidity, as indeed one
expects an overall dependence on the $|y_1-y_2|$ difference.

Confining our attention to
a rapidity interval within the central
region (where single spectra are approximately constant)
we finally get 
\begin{equation}\label{eq:corre}
C_2(y_1,y_2)= (\bar{\rho}^{(1)})^2 \ D_c^2\ +\ \langle N_c \rangle\
C_2^{(1)}(y_1,y_2) 
\end{equation}
where $\bar{\rho}_1^{(1)}$ stands for the average (charged) particle density 
from a single cluster decay, obeying the relation
$\bar{\rho}= \langle N_c \rangle \cdot \bar{\rho}^{(1)}$ with
$\bar{\rho}$ denoting the average one-particle density
in $pp$ collisions.

From Eqs.(\ref{eq:C2sl}) and (\ref{eq:corre})
one can identify 
\begin{equation}
C_2^{LR}= D_c^2\ (\bar{\rho}^{(1)})^2;\ \
C_2^{SR}= \langle N_c \rangle\ C_2^{(1)}(y_1,y_2)
\end{equation}

\subsection{Factorial moments}

From Eq.(\ref{eq:corre}) it is easy to obtain
 \begin{equation}\label{eq:F2twostep}
F_2^{(2)}=F_2^{(c)}+\frac{F_2^{(1)}}{\langle N_c \rangle}
\end{equation}
where $F_2^{(c)}=\langle N_c(N_c-1) \rangle/\langle N_c \rangle^2$ and 
$F_2^{(1)}=\langle n_1(n_1-1) \rangle/\langle n_1 \rangle^2$, and 
$\langle n_1 \rangle$ stands for the average particle multiplicity per single cluster decay.

In the Appendix we present several expressions for $F_{3,4}^{(j)}$ leaving
higher rank (i.e. $q \ge 5$) $F_q^{(j)}$ moments for a forthcoming longer paper. Let us 
also mention that $F_2^{(j)}$ often determines the values of higher rank moments in different approaches, 
and this is, in fact, the method to be followed in this work.

\section{Three-step cascade: multiple strings}

Let us develop to some extent some expressions for correlations and scaled factorial moments for
a 3-step partonic cascade.

\subsection{Two-particle correlations}
 
As in the previous 2-step cascade, by integration over rapidities of intermediate sources of particles in the
central rapidity region, 
the SR part of the 2-particle correlation function 
can be identified now as:
\begin{equation}\label{eq:C2SR}
C_2^{SR}(y_1,y_2)=\langle\ N_c \rangle\ C_2^{(1)}(y_1,y_2)
\end{equation}
The LG part reads:
\begin{equation}\label{eq:C2LR}
C_2^{LR}/(\bar{\rho}^{(1)})^2\ =\ 
\langle N_c^s \rangle^2\ D_s^2\ +\ 
\langle N_s \rangle\ D_c^2
\end{equation}
with $D_s^2=\langle N_s^2 \rangle -\langle N_s \rangle^2$. 
When setting $\langle N_s \rangle =1$ and $D_s^2=0$ 
Eq.(\ref{eq:corre}) is quickly recovered. Also note that when passing from a 2-step
to a 3-step scenario the LR contribution tends to increase.

\begin{figure}[ht!]
\begin{center}
\includegraphics[scale=0.6]{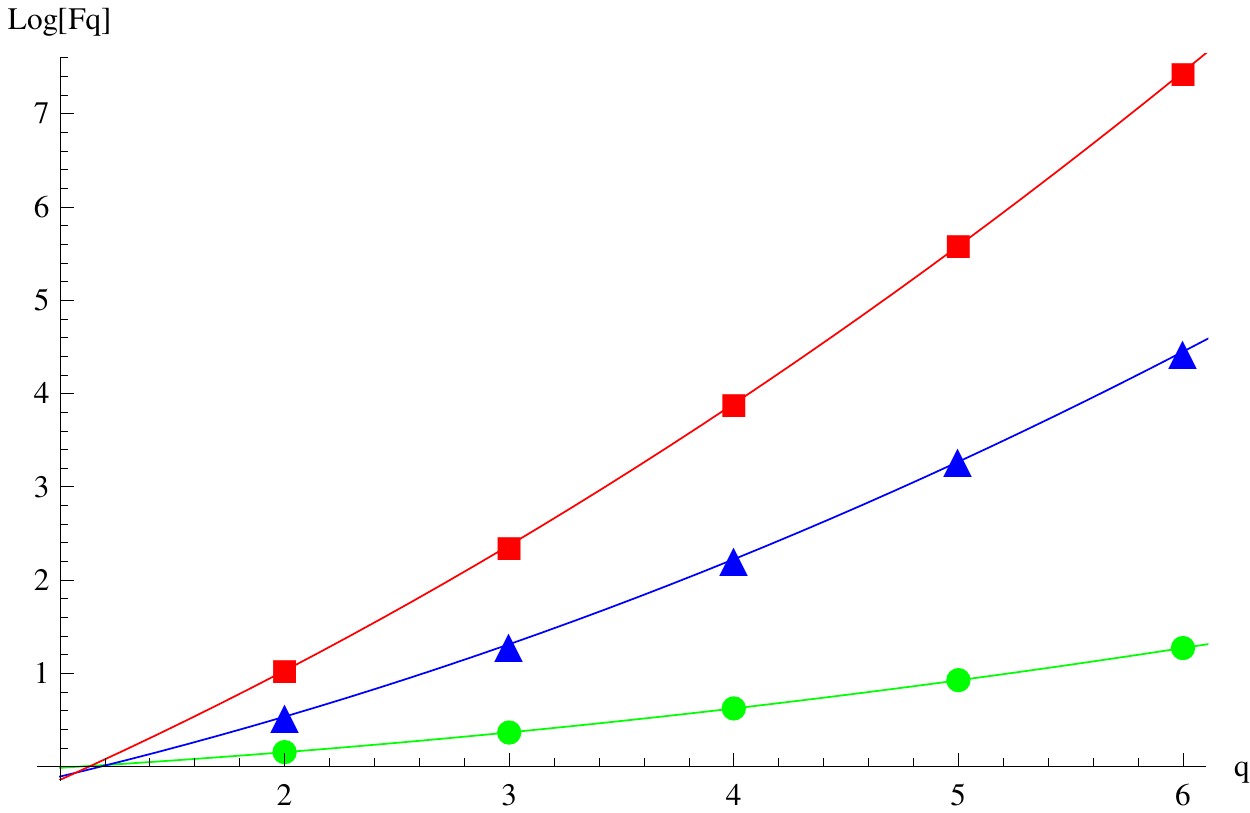}
\end{center}
\caption{Expectations for the normalized moments $\ln{F_q^{(2,3,4)}}$ versus
the rank $q$ according to: Green circles: 2-step cascade with $\langle N_c \rangle=16$; 
Blue triangles: 3-step cascade with $\langle N_s \rangle=2$, $\langle N_c \rangle=16$:
Red squares: 4-step cascade with $\langle N_h \rangle=1$, $\langle N_s \rangle=2$, 
$\langle N_c \rangle=16$. All $F_q^{(h,s,c,1)} \simeq 1$. Solid 
lines correspond to parabolic fits to points.}
\end{figure}

\subsection{Factorial moments}
\vspace{0.1cm}
 
\begin{equation}\label{eq:F2threestep}
F_2^{(3)}=F_2^{(s)}+\frac{F_2^{(c)}}{\langle N_s \rangle}+
\frac{F_2^{(1)}}{\langle N_c \rangle}
\end{equation}
where $\langle N_c \rangle=\langle N_s \rangle \cdot \langle N_c^s \rangle$; 
the scaled moment of the
distribution of fragmenting strings and clusters are here respectively defined as
$F_2^{(s)}=\langle N_s(N_s-1) \rangle/\langle N_s \rangle^2$ and  
$F_2^{(c)}=\langle N_c^s(N_c^s-1) \rangle/\langle N_c^s \rangle^2$.

\section{Four-step cascade: hidden sources}

As previously argued, we model a possible HS contribution to the standard parton cascade
as a new stage yielding a 4-step scenario, thereby modifying rapidity correlations and factorial moments
as we study below focusing on the central rapidity region of $pp$ collisions.

\subsection{Two-particle correlations}
The SR part reads again
\[
C_2^{SR}(y_1,y_2)\ =\langle\ N_c \rangle\ C_2^{(1)}(y_1,y_2)
\]
while the LR part becomes:
\[
C_2^{LR}/(\bar{\rho}^{(1)})^2=
\langle N_s \rangle^2\ D_h^2+
\langle N_h \rangle\ \langle N_c^s \rangle^2 D_s^2+ 
\langle N_s \rangle\ D_c^2
\]
where now $\langle N_s \rangle = \langle N_h \rangle\cdot\langle N_s^h \rangle$,
$\langle N_c \rangle=\langle N_s \rangle\cdot \langle N_c^s \rangle$, and 
$D_h$ denotes the dispersion of the hidden source distribution.

\subsection{Factorial moments}

\begin{equation}\label{eq:F2fourstep}
F_2^{(4)}=F_2^{(h)}+\frac{F_2^{(s)}}{\langle N_h \rangle}+\frac{F_2^{(c)}}{\langle N_s \rangle}+
\frac{F_2^{(1)}}{\langle N_c \rangle}
\end{equation}
with $F_2^{(s)}=\langle N_s^h(N_s^h-1) \rangle/\langle N_s^h \rangle^2$ and
$F_2^{(h)}=\langle N_h(N_h-1) \rangle/\langle N_h \rangle^2$.

\begin{figure}[ht!]
\begin{center}
\includegraphics[scale=0.6]{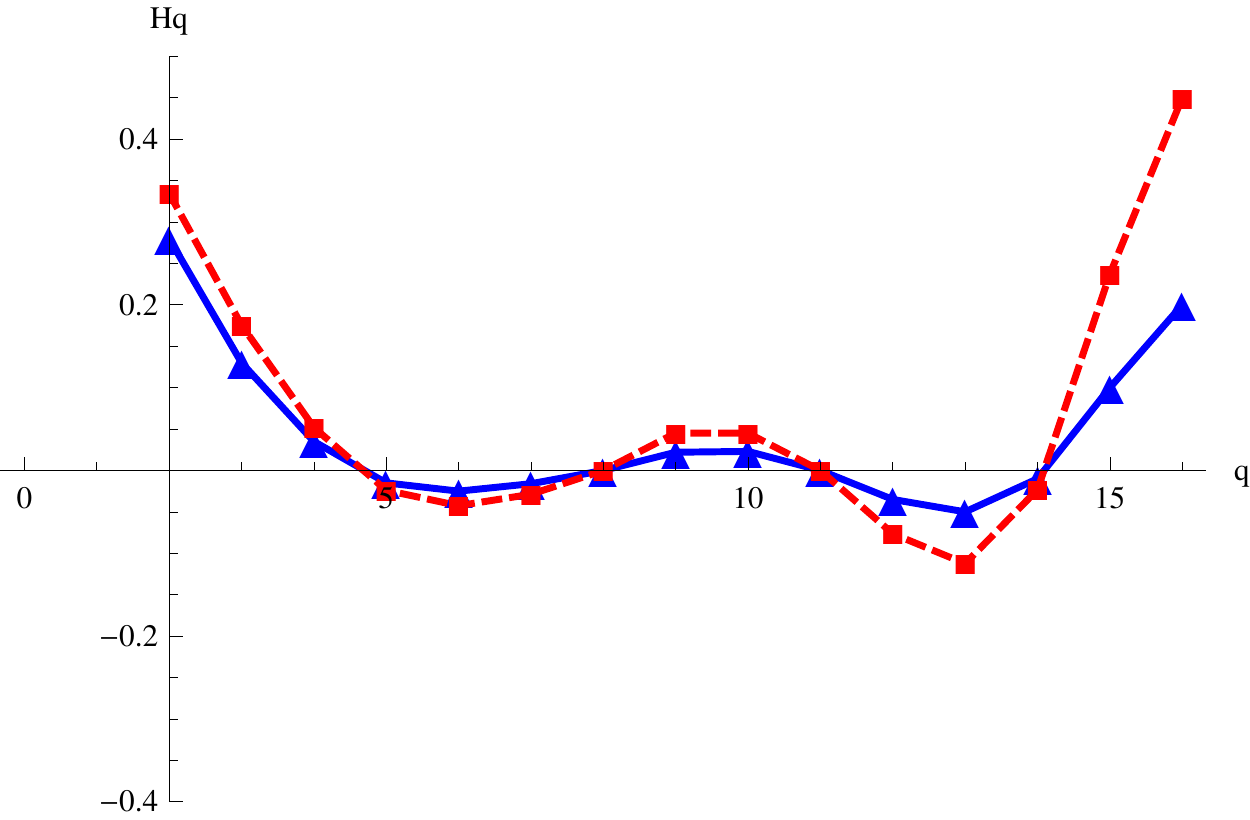}
\includegraphics[scale=0.6]{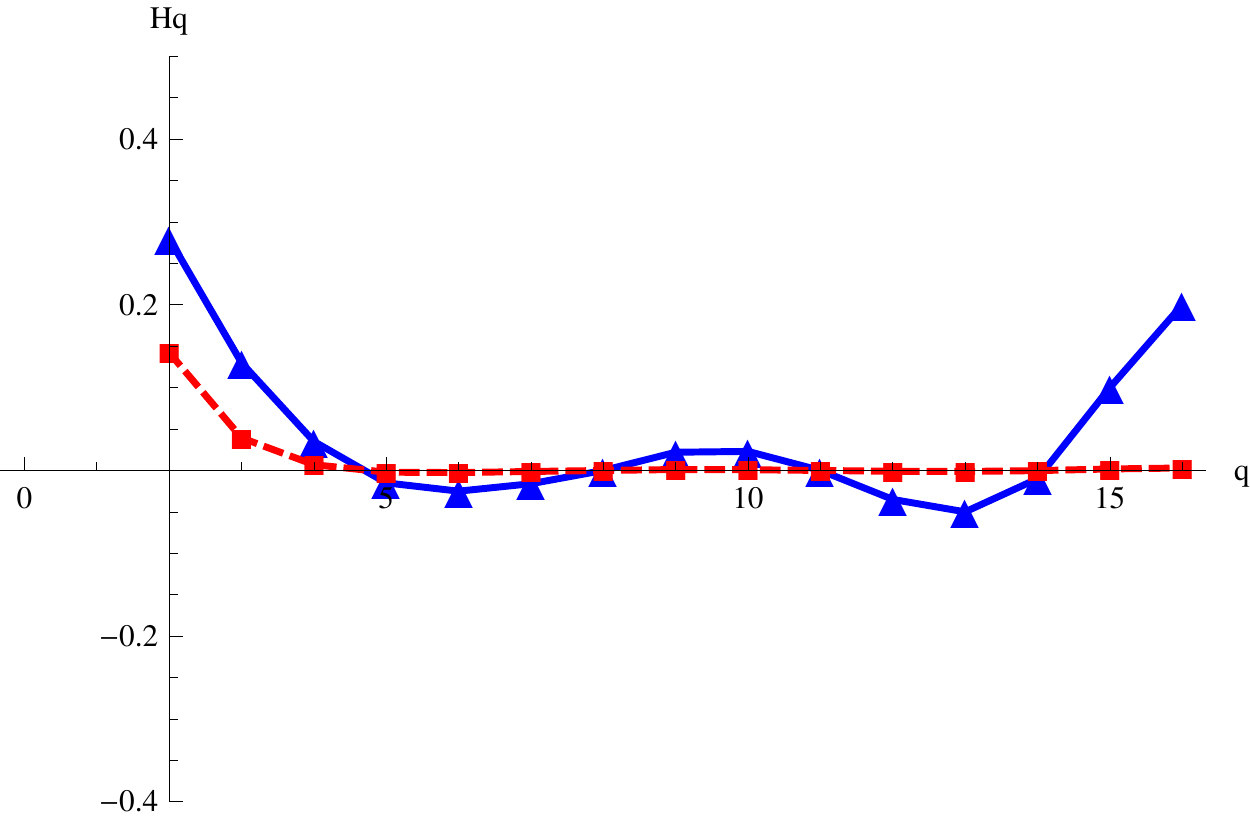}
\end{center}
\caption{Expectations for the ratios $H_q^{(3,4)}=K_q^{(3,4)}/F_q^{(3,4)}$. 
Upper panel: $\langle N_h \rangle =1$; Lower panel: $\langle N_h \rangle =8$; 
Blue triangles and solid line: conventional parton cascade according to
Ref.\cite{Dremin:2004ts}. Red squares and dashed line: hidden source(s) 
on top of the parton cascade.}
\end{figure}

\section{HS-cascade versus a conventional cascade}

In this work we advocate that a new stage at the onset of the partonic cascade can
appreciably change the multiplicity distribution of final-state particles
in pp collisions. To this aim we implement such a new stage in the conventional cascade
described as a 3-step (multi-string) process thereby becoming  a 4-step cascade.

Let us remark that, keeping fixed the number of steps of the parton cascade,  
a higher number of sources (e.g. $\langle N_c \rangle$) should lead to smaller 
$F_q^{(2,3,4)}$ moments, in accordance with the dilution effect \cite{Alexander:2000ux,Alexander:2000}. 

However, increasing the number of steps in the cascade
leads quite generally (except for large values of $\langle N_h \rangle$)
to an increase of the $F_q^{(2,3,4)}$ moments, as can be seen in Fig.1 where the green (triangle), blue (square) and red (circle)
points represent 2-, 3- and 4-step cascades, respectively. 
It is worth pointing out that larger values of 
$\langle N_h \rangle$ would considerably lower the values of $F_q^{(4)}$, even below the 2-step
and 3-step cascades.

\subsection{Analysis of $H_q^{(3,4)}$ oscillations}

Let us get started by wrtiting the expression for the $H_q$ ratios corresponding to a NBD
\begin{equation}\label{eq:bnd}
H_q^{(3,4)}= \frac{\Gamma(q)\ \Gamma(k_{eff}^{(3,4)}+1)}{\Gamma(k_{eff}^{(3,4)}+q)}
=k_{eff}^{(3,4)}\ B(q,k_{eff}^{(3,4)})
\end{equation}
where an {\em effective} $k_{eff}^{(3,4)}$ parameter is obtained from 
\begin{equation}
\frac{1}{k_{eff}^{(3,4)}}=F_2^{(3,4)}-1
\end{equation}

This approach relies on the fact that the resulting final-state
multiplicity distribution
is itself a NBD whose $k$ parameter determines the $H_q^{(3,4)}$ ratios. 
In a forthcoming paper, we will calculate $H_q^{(3,4)}$ from specific expressions of
$F_q^{(3,4)}$; some of which are shown in the appendix.

Assuming Poisson-like distributions at all intermediate steps of the cascade ($F_q^{(h,s,c,1)} \simeq 1$), one 
gets for the 2-step scenario $k_{eff}^{(2)} \lesssim \langle N_c \rangle$.\\

For the 3-step scenario one obtains
\begin{equation}\label{eq:keff3}
k_{eff}^{(3)}=\frac{\langle N_s \rangle \langle N_c \rangle}{\langle N_c \rangle+1}
\end{equation} 
If $\langle N_c \rangle >> 1$, $k_{eff}^{(3)}\ \lesssim\ \langle N_s \rangle$.\\

Finally we get for the 4-step scenario
\begin{equation}\label{eq:keff4}
k_{eff}^{(4)}=
\frac{\langle N_h \rangle \langle N_s \rangle \langle N_c \rangle}
{\langle N_h \rangle \langle N_s \rangle + \langle N_h \rangle 
\langle N_c \rangle + \langle N_s \rangle \langle N_c \rangle +1} 
\end{equation}
If besides
$\langle N_h \rangle << \langle N_s \rangle < \langle N_c \rangle$ then
$k_{eff}^{(4)} \lesssim \langle N_h \rangle$.

\subsection{Results}

Let us define a rescaling factor $r_q$ for each value of $q$ as the ratio
\begin{equation}\label{eq:rq}
r_q= \frac{H_q^{(4)}}{H_q^{(3)}}
\end{equation}
which should allow us to calculate $H_q^{(4)}$ 
from the a 3-step conventional cascade that we assume to coincide with the
expectations for $pp$ collisions at the LHC (14 TeV) 
obtained in Ref.\cite{Dremin:2004ts}.

Thus, squares and the dashed line (in red) of Fig.2 (upper and lower panels)  
correspond to a hidden source in a 4-step cascade, with $N_h=1$, $N_s^h=2$, 
$N_c^s=8$, and $N_h=8$, $N_s^h=2$, $N_c^s=8$, respectively. Triangles and
the solid line (in blue) correspond to the points from the
above-mentioned Ref.\cite{Dremin:2004ts}.

Admittedly, this is a rough approach in order to
compare $H_q^{(3,4)}$ oscillations, i.e. multiparticle production
{\em with} and {\em without} a HS. As
already advocated in this paper, the key idea is that a (statistically significant) 
deviation one from each other might prove to be useful to unravel
a possible signal of a hidden sector participating in the parton cascade.

\section{Summary and final remarks}

In this work we advocate that a new stage of matter (stemming from a hidden sector beyond the SM) on top
of the partonic cascade leading to multiparticle final states in $pp$ collisions at the LHC can have
an influence on rapidity correlations and factorial moments (and their ratios) of multiplicity distributions.

On the one hand, as a general trend one should expect that a HS yields longer (in rapidity) and stronger 
2-particle correlations among emitted charged particles. 

On the other hand, depending on the number of hidden sources two different behaviours of the oscillation 
pattern of $H_q^{(4)}$ moments can be distinguished:

\begin{itemize}

\item For a small number of hidden sources (e.g. $\langle N_h \rangle=1$), the oscillation
amplitudes become considerably larger than in a conventional cascade as obtained in \cite{Dremin:2004ts}.
See figure 2 (upper panel). 

\item For a larger number of hidden sources (e.g. $\langle N_h \rangle=8$), the oscillation
amplitudes are considerably smaller than in a conventional cascade as obtained in \cite{Dremin:2004ts}.
See figure 2 (lower panel). Let us note that
the crossing point shifts to a smaller $q$ value for higher $\langle N_h \rangle$.

\end{itemize}

Let us point out that that there could be combinations of parameters such that $H_q^{(3,4)}$ 
oscillations would become practically indistinguishable from each other.

Moreover, both (conventional and
HS) process will be present together in the collected sample of events, although specific cuts 
(such as high multiplicity, flavour tagging, etc) should be applied on events to enrich the possible
new physics contribution.  

Another caveat is in order. The behaviour of such oscillations is very sensitive, e.g., 
to multiplicity cuts on events \cite{Dremin:2000ep}. Therefore, to avoid a bias 
one should be careful to compare samples at different energies 
with the same cuts applied on events.

\section{Acknowledgments}

M.A. Sanchis-Lozano wants to thank the organizers of ICHEP for the excellent organization
in all respects. This work has been partially supported by MINECO under grant FPA-23596
and Generalitat Valenciana under grant PROMETEOII/2014/049.

\appendix

\section{Factorial moments for 2-, 3- and 4-step scenarios}

In this appendix we only show expressions for factorial moments $F_3^{(j)}$ and $F_4^{(j)}$
(where $j=2,3,4$ denotes 2-, 3- or 4-step scenarios, respectively) and leave higher rank factorial 
moments for a long paper in preparation.\\

\noindent
{\bf 2-step cascade}
\[
F_3^{(2)}\ =\ F_3^{(c)}\ +\ 3\ \frac{F_2^{(c)}}{\langle N_c \rangle}F_2^{(1)}\ 
+\ \frac{F_3^{(1)}}{\langle N_c \rangle^2}
\]

\[
F_4^{(2)}=F_4^{(c)}+\frac{F_4^{(1)}}{\langle N_c \rangle^3}+
4\ \frac{F_2^{(c)}F_3^{(1)}}{\langle N_c \rangle^2}+
\]
\[
6\ \frac{F_3^{(c)}F_2^{(1)}}
{\langle N_c \rangle}+3\ \frac{F_2^{(c)}F_2^{(1)2}}{\langle N_c \rangle^2}
\]

\noindent
{\bf 3-step cascade}
\[
F_3^{(3)}=F_3^{(s)}+\frac{F_3^{(c)}}{\langle N_s \rangle^2}+
3\ \biggl[\frac{F_2^{(s)}F_2^{(c)}}{\langle N_s \rangle}+\frac{F_2^{(s)}
F_2^{(1)}}{\langle N_c \rangle}+
\]
\[
\frac{F_2^{(c)}F_2^{(1)}}{\langle N_s \rangle \langle N_c \rangle}
\biggr]+ 
\frac{F_3^{(1)}}{\langle N_c \rangle^2}
\]

\[
F_4^{(3)}=F_4^{(s)}+\frac{F_4^{(c)}}{\langle N_s \rangle^3}+ 
\frac{F_4^{(1)}}{\langle N_c \rangle^3}+ 
\]
\[
4\ \biggl[\frac{F_2^{(s)}F_3^{(c)}}{\langle N_s \rangle^2}+
\frac{F_2^{(s)}F_3^{(1)}}{\langle N_c \rangle^2}+
\frac{F_2^{(c)}F_3^{(1)}}{\langle N_s \rangle \langle N_c \rangle^2}
\biggr]+ 
\]
\[
6\ \biggl[
\frac{F_3^{(s)}F_2^{(c)}}{\langle N_s \rangle}+
\frac{F_3^{(s)}F_2^{(1)}}{\langle N_c \rangle}+
\frac{F_3^{(c)}F_2^{(1)}}{\langle N_s \rangle^2 \langle N_c \rangle}
\biggr]+ 
\]
\[
3\ \biggl[\frac{F_2^{(s)}F_2^{(c)2}}{\langle N_s \rangle^2}+
\frac{F_2^{(s)}F_2^{(1)2}}{\langle N_c \rangle^2}+
\frac{F_2^{(c)}F_2^{(1)2}}{\langle N_s \rangle \langle N_c \rangle^2}\biggr]+ 
\]
\[
18\ \frac{F_2^{(s)}F_2^{(c)}F_2^{(1)}}{\langle N_s \rangle \langle N_c \rangle}
\]

\noindent
{\bf 4-step cascade}
\[
F_3^{(4)}=F_3^{(h)}+\frac{F_3^{(s)}}{\langle N_h \rangle^2}+
\frac{F_3^{(c)}}{\langle N_s \rangle^2}+
\frac{F_3^{(1)}}{\langle N_c \rangle^2}+ 
\]
\[
3 \biggl[\frac{F_2^{(s)}F_2^{(1)}}{\langle N_h \rangle \langle N_c \rangle}+
\frac{F_2^{(s)}F_2^{(c)}}{\langle N_h \rangle \langle N_c \rangle}+
\frac{F_2^{(c)}F_2^{(1)}}{\langle N_s \rangle \langle N_c \rangle}+
\]
\[
\frac{F_2^{(h)}F_2^{(1)}}{\langle N_c \rangle}+
\frac{F_2^{(h)}F_2^{(c)}}{\langle N_s \rangle}+ 
\frac{F_2^{(h)}F_2^{(s)}}{\langle N_h \rangle}
\biggr]
\]

\[
F_4^{(4)} =  F_4^{(h)}+\frac{F_4^{(s)}}{\langle N_h \rangle^3}+
\frac{F_4^{(c)}}{\langle N_s \rangle^3}+
\frac{F_4^{(1)}}{\langle N_c \rangle^3}+
\]
\[
4\ \biggl[
\frac{F_2^{(s)}F_3^{(c)}}{\langle N_h \rangle\langle N_s \rangle^2}+
\frac{F_2^{(s)}F_3^{(1)}}{\langle N_h \rangle\langle N_c \rangle^2}+
\frac{F_2^{(c)}F_3^{(1)}}{\langle N_s \rangle\langle N_c \rangle^2}+
\]
\[ 
\frac{F_2^{(h)}F_3^{(s)}}{\langle N_h \rangle^2}+
\frac{F_2^{(h)}F_3^{(c)}}{\langle N_s \rangle^2}+
\frac{F_2^{(h)}F_3^{(1)}}{\langle N_c \rangle^2}
\biggr]+
\]
\[
6\ \biggl[
\frac{F_3^{(s)}F_2^{(c)}}{\langle N_h \rangle^2
\langle N_s \rangle}+
\frac{F_3^{(s)}F_2^{(1)}}{\langle N_h \rangle^2
\langle N_c \rangle}+
\frac{F_3^{(c)}F_2^{(1)}}{\langle N_s \rangle^2
\langle N_c \rangle}+
\]
\[
\frac{F_3^{(h)}F_2^{(c)}}{\langle N_s \rangle}+
\frac{F_3^{(h)}F_2^{(1)}}{\langle N_c \rangle}+
\frac{F_3^{(h)}F_2^{(s)}}{\langle N_h \rangle}\biggr]+
\]
\[
3\ \biggl[
\frac{F_2^{(s)}F_2^{(1)2}}{\langle N_h \rangle
\langle N_c \rangle^2}+
\frac{F_2^{(s)}F_2^{(1)2}}{\langle N_h \rangle
\langle N_s \rangle^2}+
\frac{F_2^{(c)}F_2^{(1)2}}{\langle N_s \rangle\langle N_c \rangle^2}+
\]
\[
\frac{F_2^{(h)}F_2^{(1)2}}{\langle N_c \rangle^2}+
\frac{F_2^{(h)}F_2^{(c)2}}{\langle N_s \rangle^2}+
\frac{F_2^{(h)}F_2^{(s)2}}{\langle N_h \rangle^2}\biggr]+
\]
\[
18\ \biggl[\frac{F_2^{(s)}F_2^{(c)}F_2^{(1)}}{\langle N_h \rangle
\langle N_s \rangle \langle N_c \rangle}+\frac{F_2^{(h)}F_2^{(c)}F_2^{(1)}}
{\langle N_s \rangle \langle N_c \rangle}+
\frac{F_2^{(h)}F_2^{(s)}F_2^{(c)}}{\langle N_h \rangle
\langle N_s \rangle}+ 
\]
\[
\frac{F_2^{(h)}F_2^{(s)}F_2^{(1)}}{\langle N_h \rangle \langle N_c \rangle}
\biggr] 
\]



%% References

%\nocite{*}
%\bibliographystyle{elsarticle-num}
%\bibliography{martin}






% \end{thebibliography}

\end{document}